\documentclass[useAMS,usenatbib]{mn2e}

\usepackage{amssymb}
\usepackage{times}
\usepackage{graphicx}
\usepackage{epstopdf}
\usepackage{subfigure}

\newcommand{\beq}{\begin{equation}}
\newcommand{\eeq}{\end{equation}}

\newcommand{\bfbeta}{\mbox{\boldmath $\beta$}}

\newcommand{\bftheta}{\mbox{\boldmath $\theta$}}

\newcommand{\bfP}{\mathbf{P}}

\def\gs{\mathrel{\lower0.6ex\hbox{$\buildrel {\textstyle >}\over{\scriptstyle \sim}$}}}
\def\ls{\mathrel{\lower0.6ex\hbox{$\buildrel {\textstyle <}\over{\scriptstyle \sim}$}}}
\newcommand{\simgt}{\lower.5ex\hbox{$\; \buildrel > \over \sim \;$}}
\newcommand{\simlt}{\lower.5ex\hbox{$\; \buildrel < \over \sim \;$}}

\newcommand{\aap}{A\&A}
\newcommand{\apj}{ApJ}
\newcommand{\apjl}{ApJ}

\newcommand{\aj}{AJ}
\newcommand{\na}{New A.}

\newcommand{\prd}{Phys. Rev. D}
\newcommand{\mnras}{MNRAS}

\begin{document}

\title[Time delay cosmography]{Hubble constant and dark energy inferred from free-form determined time delay distances}
\author[M. Sereno and D. Paraficz]{
Mauro Sereno$^{1,2}$\thanks{E-mail: mauro.sereno@polito.it (MS)}, Danuta Paraficz$^{3}$
\\
$^1$Dipartimento di Scienza Applicata e Tecnologia, Politecnico di Torino, corso Duca degli Abruzzi 24, I-10129 Torino, Italia\\
$^2$INFN, Sezione di Torino, via Pietro Giuria 1, I-10125, Torino, Italia\\
$^3$Aix Marseille UniversitŽ, CNRS, LAM (Laboratoire d'Astrophysique de Marseille) UMR 7326, 13388, Marseille, France
}


\maketitle

\begin{abstract}
Time delays between multiple images of lensed sources can probe the geometry of the universe. We propose a novel method based on free-form modelling of gravitational lenses to estimate time-delay distances and, in turn, cosmological parameters. This approach does not suffer from the degeneracy between the steepness of the profile and the cosmological parameters. We apply the method to 18 systems having time delay measurements and find $H_0=69\pm6\mathrm{(stat.)}\pm4\mathrm{(syst.)}~\mathrm{km~s^{-1}Mpc^{-1}}$. In combination with WMAP9, the constraints on dark energy are $\Omega_w=0.68\pm0.05$ and $w=-0.86\pm0.17$ in a flat model with constant equation-of-state.
\end{abstract}


\section{Introduction}
Multiple images of lensed sources take different times to complete their travel and probe the potential of the deflector as well as the geometry of the universe. This has been long known and proposed as a test to measure the Hubble constant, $H_0$, and the dark energy \citep{ref64,koc04,lin11,tre+al13}. The difference in arrival time $\Delta t$ is given by 
\beq
\Delta t \propto D_{\Delta t} \Delta \phi,
\eeq
where the Fermat potential $\Delta \phi$ contains all of the dependence on the mass distribution whereas the time delay distance $D_{\Delta t}(\propto H_0^{-1})$ depends only on the cosmology. Even if very appealing on the theoretical side, cosmographic tools based on time delay measurements have been hampered by severe model dependent uncertainties, regardless of the quality of the lensing data. Due to intrinsic degeneracies, observed image positions or rings, magnification ratios, and time delays can be reproduced by different mass models.

In order to address this problem, two main approaches have been proposed. Within the first approach, degeneracies are broken by complementing lensing data with additional information that constrains the lens mass profile, such as the measurements of the stellar velocity dispersion or the mass distribution along the line of sight \citep{suy+al13}. This method can yield time-delay distances measured to $\sim 6$ per cent precision \citep{suy+al10,suy+al13}.

This formal precision relies on the assumption that the mass profile is well determined. The uncertainty in the measurement of the time-delay distances has to be realistic with respect to various choices of parametric models. Reassuringly, a simple modelling such as a global power law may provide adequate results even in complex systems. \citet{suy+al10} found that a pixelated reconstruction of the potential of the galaxy lensing B~1608+656 does not show significant corrections from a global power law. \citet{suy+al13b} found similar results for RXJ~1131-1231 by using flexible gravitational lens models with baryonic and dark matter components.

However, mass-sheet like transformations leave all lensing observables exactly invariant, except the product of time delay and cosmological distances. No matter whether the environment of the lens is very well characterised, this degeneracy can still manifest through the correlation between the inner slope of the mass profile and the cosmological parameters.  \citet{sc+sl13} argued that notwithstanding stellar kinematics measurements degeneracies between parametrized models could imply uncertainties on $H_0$ of the order of 10-20 per cent.

An alternative strategy for dealing with the problem of non uniqueness is to search through a large ensemble of models that can all reproduce the observations. This can be performed with free-form model reconstructions that allow a wide range of solutions \citep{sa+wi04,col08}. Within this approach, the time-delay distance is determined with a larger statistical uncertainty, $\gs 30$ per cent \citep{sah+al06,ogu07}. However, even if the mass-sheet degeneracy still plays a role, results are not biased by the degeneracy between with the density slope and the cosmological parameters.

The assessment of the real uncertainty is crucial in view of the potential of time-delay distances as cosmological probes. This estimate is appealing since it is based on simple geometry and well-tested physics and produces a direct measurement of a cosmological distance \citep{tre+al13}. It can directly measure the Hubble constant and it is complementary with cosmic microwave background (CMB) measurements to constrain the dark energy equation-of-state and its evolution \citep{lin11}.

Wide field imaging surveys are likely to discover thousands of lensed quasars within the next decade. If a single lens can actually provide a distance measurement with 5 per cent accuracy, the targeted study of one hundred of these systems can result in substantial gains in the dark energy figure of merit \citep{lin11,tre+al13}. In the next decade, LSST (Large Synoptic Survey Telescope) and Euclid will enlarge the sample to thousands of systems and detailed cosmographic analyses will be effective even with the more conservative estimate of uncertainties of 30-40 per cent per system.

Here, we study what can be done with the gravitational lens systems observed up to date. We develop a method to measure the Hubble constant and to characterise the dark energy based on free-form modelling and apply it to time-delay measurements already available. Similar approaches were successfully used in the past to constrain $H_0$ under the assumption that dark matter density and dark energy were known \citep{sah+al06,col08,pa+hj10}. We extend the method to infer the time-delay distance and, in turn, the Hubble constant together with other cosmological parameters.

The paper is organised as follows. In Section~\ref{sec_free} we summarise the basics of the free-form modelling we use and how to apply it to the measurement of the time delay distance. Section~\ref{sec_sample} presents the lens sample. The statistical analysis of the data and the derivation of the cosmological parameters is described in section~\ref{sec_analysis}. Finally, section~\ref{sec_conc} is devoted to conclusions.

\section{Free-form analysis}
\label{sec_free}

Our analysis requires reliable free-form modelling of mass maps of gravitational lenses, as can be obtained with the PixeLens formalism \citep{sa+wi04}. PixeLens has been already successfully tested with simulations and applied to real data \citep{sa+wi04,sah+al06,col08,pa+hj10,se+zi12} and we refer to the quoted papers for details. Here, we briefly summarise the main features behind the determination of the time delay distance. PixeLens generates an ensemble of lens models that exactly fit the image positions and the time delays. Each model is made up of the set of the $n$ convergences $\kappa_n$ of the pixels which discretise the lens surface mass distribution, the source positions $\bfbeta$, the time delay distance $D_{\Delta t}$, and optionally the external shear. 

The time delay between two images at positions $\bftheta_i$ and $\bftheta_j$ can be expressed as
\beq
\tau (\bftheta_i)-\tau (\bftheta_j) = \frac {c \Delta t_{ij}}{D_{\Delta t}},
\eeq
where $\Delta t_{ij}$ is the measured time delay and $\tau$ is the dimensionless arrival time depending on the lens properties. The time-delay distance is defined as
\beq
D_{\Delta t}=(1+z_\mathrm{d})\frac{D_\mathrm{d} D_\mathrm{s}}{D_\mathrm{ds}},
\eeq
where $z_\mathrm{d}$ is the deflector redshift and $D_\mathrm{d}$, $D_\mathrm{s}$ and $D_\mathrm{ds}$ are the angular diameter distances from the observer to the lens, from the observer to the source and from the lens to the source, respectively. For a pixelated lens model, the arrival time is \citep{sa+wi04},
\beq
\tau(\bftheta)=\frac{1}{2}|\bftheta|^2-\bftheta \cdot \bfbeta -\sum_n \kappa_n Q_n (\bftheta), 
\eeq
where $Q_n$ is given in \citet{col08}. An external shear can be linearly added to the arrival time. Since all variables have to appear linearly in the equations to be managed by PixeLens, the actual variable in the lens model is $D_{\Delta t}^{-1}$ instead of $D_{\Delta t}$.

Time delays depend on cosmological parameters in two main ways. The main dependence comes from the time-delay distance. The time-delay is proportional to $D_{\Delta t}$ and provides a direct measurement of a cosmological distance.

The second dependence comes from the dimensionless convergence $\kappa$, which is proportional to $D_\mathrm{ds}/D_{s}$. For a single source redshift, as for most of the lenses with measured time-delays, this factor is completely degenerate with the mass normalisation and cannot be directly inferred without a mass estimate within the Einstein radius independent of lensing. We can consider this dependence as a secondary effect.

PixeLens uses a Bayesian approach. Discretized probability distributions for the parameters are obtained by collecting the parameters values of the model ensamble. In particular, the probability distribution of time-delay distance, which contains the main cosmological dependence, is obtained after marginalisation over the pixel convergences. In this way, we also de facto get rid of the dependence on the distance ratio.

The dependence on $D_\mathrm{ds}/D_{s}$ can be instead exploited to constrain dark energy with galaxy cluster lenses characterised by image systems at multiple redshifts \citep{ser02,se+lo04,sou+al04,gi+na09,jul+al10}. In the framework of free-form modelling, one way to exploit this dependence is to take the volume of the solution space as a tracer of the probability of the underlying cosmological assumption \citep{lub+al13}.

In previous analyses with PixeLens the cosmological parameters other than $H_0$ were fixed \citep{sah+al06,col08,pa+hj10}. In this way, the proportionality factors between the time-delay distances and $H_0$ were assumed to be known. All lenses could be analysed at once. Instead of one unknown time-delay distance for each lens, there was just a single unknown variable, i.e., the Hubble constant, which was shared by all of the lenses. $H_0$ was then allowed to vary from ensemble model to model but not from lens to lens within a single model. The final output was the posterior probability function for $H_0$.

Here, we study dark energy as well as the Hubble constant. We have to estimate the time-delay distance for each system. We then model each lens separately and end up with an ensemble of distributions of time-delay distances (measured at different redshifts).

\section{Lenses}
\label{sec_sample}

\begin{table}
\centering
\caption{List of lenses. Number of images are reported in col.~2. References for time delays are quoted in col.~3: (1) \citet{eu+ma11}; (2) \citet{cou+al11}; (3) \citet{goi+al08}; (4) \citet{ull+al06}; (5) \citet{osc+al01}; (6) \citet{foh+al08}; (7) \citet{tew+al12}; (8) \citet{eul+al13}; (9) \citet{vui+al07}; (10) \citet{lov+al98}; (11) \citet{vui+al08}.
}
\label{tab_sample}
\begin{tabular}[c]{lcccl}
        \hline
        \noalign{\smallskip}
        name &	$N_\mathrm{img}$	& ref.    \\
        \noalign{\smallskip}
        \hline
JVAS B0218+357	&	2	&	(1)	\\
HE 0435-1223		&	4	&	(2)	\\
SBS 0909+532		&	2	&	(3,4)	\\
RX J0911+0551	&	4	&	(1)	\\
FBQS J0951+2635	&	2	&	(1)	\\
Q J0957+561		&	2, 2	&	(5)	\\
SDSS J1004+4112	&	4	&	(6)	\\
HE 1104-1805		&	2	&	(1)	\\
PG 1115+080		&	4	&	(1)	\\
RX J1131-1231	&	4	&	(7)	\\
SDSS J1206+4332	&	2	&	(8)	\\
SBS 1520+530		&	2	&	(1)	\\
CLASS B1600+434	&	2	&	(1)	\\
CLASS B1608+656	&	4	&	(1)	\\
SBS J1650+4251	&	2	&	(9)	\\
PKS 1830-211		&	2	&	(10)	\\
WFI J2033-4723	&	4	&	(11)	\\
HE 2149-2745		&	2	&	(1)	\\
       \hline
       \end{tabular}
\end{table}

Our sample consists of 18 lenses with both measured time delays and spectroscopic determination of deflector and source redshifts, see Table~\ref{tab_sample}. There are a few more lenses with time delay measurements which we did not consider in the analysis. We excluded B~1422+231, whose time-delay measurement is uncertain \citep{eu+ma11}, and HS~2209+1914 \citep{eul+al13} and H~1413+117 \citep{goi+al10}, which lack spectroscopic measurements of $z_\mathrm{d}$. Galaxy cluster SDSS~J1029+2623 \citep{ogu+al08,foh+al13} requires a better understanding of its lensing features before the measured time delay can be safely used for cosmography.

Systems were modelled following \citet{pa+hj10}, which we refer to for details. Briefly, the mass maps of the doubly imaged quasars were required to have $180\deg$ rotation symmetry. Lenses with quadruply imaged systems were modelled as asymmetric distribution if they are known to be irregular. A constant external shear is added for the lenses where the morphology shows evidence of external distortion. 

Finally, in addition to the main lens we have also included all the galaxies that might contribute to the lensing. Extra lenses are added whenever one or more galaxies are visible in the field and when their redshift is similar to the main lens. The complete list of additional galaxies can be found in \citet[ table 2]{pa+hj10}.

Extra galaxies are modelled as point masses at the corresponding pixel location. Point masses are not meant as a realistic mass profiles for nearby galaxies. They just allows local spikes in the pixellated mass map which can override the nearest neighbour constraint implemented in PixeLens. In fact, PixeLens does not model the main lens and the extra galaxies separately. The mass map accounts for the total projected density.

The inclusion of extra lenses may also break the global $180\deg$ rotation symmetry, which continues to hold only for the main lenses of the doubly imaged quasars.

PixeLens employs some very mild priors on the mass distribution. A detailed discussion can be found in \citet{col08}. These priors are meant to preserve the positivity and the smoothness of the lens density. Even if they play an important role in the sampling strategy, they can not drive the derived properties of the lens, which are determined by the data.

The mass sheet degeneracy still affects each mass map obtained by PixeLens. In principle, PixeLens does not make any requirement on the mean density and a model in the ensemble might be just the mass-sheet transformation of a companion model. The final ensemble should then partially take into account the degeneracy. On the other hand, the sampling strategy of PixeLens prefers mass density profiles whose convergence rapidly goes to zero outside the region of the multiple images, which is similar to fixing the external value of the convergence.

A simple strategy to circumvent this problem is rescaling the distribution of time-delay distances obtained without any assumption on the mean convergence, $D_{\Delta t}(\lambda=0)$, by $D_{\Delta t}(\lambda)=D_{\Delta t}(\lambda=0)/(1-\lambda) $. We will adopt this approach in Section~\ref{sec_syst} to test some systematic effects.

The minimal approach to model lenses described above proved very efficient in modelling either simulated or real lens systems \citep{sah+al06,pa+hj10}. We generated an ensemble of 250 models for each lens. With respect to \citet{pa+hj10}, we updated the measurement of the time delays, see Table~\ref{tab_sample}.

The positions of the observed images and the redshifts of the source and lens are accurately measured. The related uncertainties are much smaller than the pixel size and the mass model variations, respectively. These uncertainties can then be ignored. Time delays between images are similarly assumed to be accurate since their uncertainties are much smaller than the range of models that reproduce the data. These assumptions have been successfully tested with synthetic lenses and numerical simulations \citep{sa+wi04,sah+al06,col08}.

\section{Analysis}
\label{sec_analysis}

\begin{table}
\centering
\caption{Time delay distances. References for the source and lens redshifts are listed in \citet[ table 2]{pa+hj10}. Quoted values are the biweight estimators of central location and width.
}
\label{tab_D_Delta_t}
\begin{tabular}[c]{lllr@{$\,\pm\,$}l}
        \hline
        \noalign{\smallskip}
	name	&	$z_\mathrm{d}$&	$z_\mathrm{s}$& \multicolumn{2}{c}{$D_{\Delta t}~[\mathrm{Mpc}]$}  \\
        \noalign{\smallskip}
        \hline
JVAS B0218+357	&	0.6847	&	0.944	&	11700	&	3600		\\
HE 0435-1223		&	0.4541	&	1.689	&	3160		&	830		\\
SBS 0909+532	&	0.83		&	1.376	&	9460		&	3500		\\
RX J0911+0551	&	0.769	&	2.8		&	3450		&	840		\\
FBQS J0951+2635	&	0.24		&	1.246	&	2110		&	810		\\
Q J0957+561		&	0.356	&	1.41		&	1650		&	530		\\
SDSS J1004+4112	&	0.68		&	1.734	&	4090		&	1400		\\
HE 1104-1805		&	0.729	&	2.319	&	3120		&	1500		\\
PG 1115+080		&	0.311	&	1.722	&	2070		&	590		\\
RX J1131-1231	&	0.295	&	0.658	&	1610		&	520		\\
SDSS J1206+4332	&	0.748	&	1.789	&	3230		&	970		\\
SBS 1520+530		&	0.761	&	1.855	&	5200		&	1900		\\
CLASS B1600+434	&	0.41		&	1.59		&	3550		&	1600		\\
CLASS B1608+656	&	0.63		&	1.394	&	6300		&	1500		\\
SBS J1650+4251	&	0.577	&	1.547	&	5600		&	2200		\\
PKS 1830-211		&	0.885	&	2.507	&	5480		&	2300		\\
WFI J2033-4723	&	0.661	&	1.66		&	5430		&	1200		\\
HE 2149-2745		&	0.495	&	2.03		&	4350		&	1900		\\
\hline
	\end{tabular}
\end{table}

\begin{table*}
\caption{Cosmological parameters. Models of universe are characterised by the different sharp priors on the cosmological parameters. Quoted values are mean and standard deviations of the posterior probability distribution. Square brackets denote parameters fixed a priori. The $\blacksquare$ symbol denotes that the parameter was left free to vary but was undetermined by the analysis.
}
\label{tab_cosm_par}
\begin{tabular}[c]{c|cr@{$\,\pm\,$}lr@{$\,\pm\,$}lr@{$\,\pm\,$}lr@{$\,\pm\,$}lr@{$\,\pm\,$}l}
        \hline
        \noalign{\smallskip}
	model	&	data sets	& \multicolumn{2}{c}{$h$} & \multicolumn{2}{c}{$\Omega_\mathrm{M}$} & \multicolumn{2}{c}{$\Omega_w$} & \multicolumn{2}{c}{$w_0$} & \multicolumn{2}{c}{$w_a$}  \\
        \noalign{\smallskip}
        \hline
OWACDM	&	time-delays			&	0.69	&	0.06	&	\multicolumn{2}{c}{$\blacksquare$}	&	\multicolumn{2}{c}{$\blacksquare$}	&	\multicolumn{2}{c}{$\blacksquare$}	&	\multicolumn{2}{c}{$\blacksquare$}\\
\hline
OWCDM	&	time-delays			&	0.68	&	0.06	&	\multicolumn{2}{c}{$\blacksquare$}	&	\multicolumn{2}{c}{$\blacksquare$}	&	\multicolumn{2}{c}{$\blacksquare$}	&	\multicolumn{2}{c}{$[0]$}\\
OWCDM	&	time-delays, WMAP9	&	0.65	&	0.06	&	0.33	&	0.06			&	0.68	&	0.06			&	$-1.01$	&	0.40		&	\multicolumn{2}{c}{$[0]$}\\
\hline
WCDM	&	time-delays			&	0.67	&	0.06	&	\multicolumn{2}{c}{$\blacksquare$}	&	\multicolumn{2}{c}{$[1-\Omega_\mathrm{M}]$}	&	\multicolumn{2}{c}{$\blacksquare$}	&	\multicolumn{2}{c}{$[0]$}\\
WCDM	&	time-delays, WMAP9	&	0.66	&	0.05	&	0.32	&	0.05			&	\multicolumn{2}{c}{$[1-\Omega_\mathrm{M}]$}	&	$-0.86$	&$0.17$	&	\multicolumn{2}{c}{$[0]$}\\
\hline
O$\Lambda$CDM	&	time-delays			&	0.65	&	0.05	&	\multicolumn{2}{c}{$\blacksquare$}	&	\multicolumn{2}{c}{$\blacksquare$}	&	\multicolumn{2}{c}{$[-1]$}	&	\multicolumn{2}{c}{$[0]$}\\
O$\Lambda$CDM	&	time-delays, WMAP9	&	0.66	&	0.03	&	0.32	&	0.04			&	0.69	&	0.04			&	\multicolumn{2}{c}{$[-1]$}	&	\multicolumn{2}{c}{$[0]$}\\
\hline
$\Lambda$CDM	&	time-delays			&	0.66	&	0.04	&	\multicolumn{2}{c}{$\blacksquare$}	&	\multicolumn{2}{c}{$[1-\Omega_\mathrm{M}]$}	&	\multicolumn{2}{c}{$[-1]$}	&	\multicolumn{2}{c}{$[0]$}\\
$\Lambda$CDM	&	time-delays, WMAP9	&	0.69	&	0.02	&	0.29	&	0.02			&	\multicolumn{2}{c}{$[1-\Omega_\mathrm{M}]$}	&	\multicolumn{2}{c}{$[-1]$}	&	\multicolumn{2}{c}{$[0]$}\\
\hline
	\end{tabular}
\end{table*}

Estimates of the time delay distances are listed in Table~\ref{tab_D_Delta_t}. Errors range from 20 to 50 per cent, which reflects the large degeneracy in mass models. Not surprisingly given the large uncertainties, the estimates for B~1608+656 and RX~J1131-1231 are in agreement with the results in \citet[ $D_{\Delta t} = 5160\pm270~\mathrm{Mpc} $]{suy+al10} and \citet[ $D_{\Delta t} = 2090\pm130~\mathrm{Mpc} $]{suy+al13}, respectively.

To constrain the cosmological parameters, we performed a standard Bayesian analysis. The likelihood of the parameters $\bfP$ is
\beq
{\cal L}( \bfP) \propto \prod_i p_i (D_{\Delta t}[z_\mathrm{d}^i,z_\mathrm{s}^i;\bfP] )
\eeq
where $p_i (D_{\Delta t})$ is the estimated time-delay distance probability distribution of the $i$-th lens. We considered flat priors for the cosmological parameters. For the Hubble constant, $0.2 \le h \le 1$, where $h$ is $H_0$ in units of $100~\mathrm{km~s}^{-1}\mathrm{Mpc}^{-1}$; for the matter density, $0\le \Omega_\mathrm{M}\le 1$; for the dark energy density, $0\le \Omega_w \le 1$; for the dark energy equation-of-state $w(a)=w_0+ (1-a)w_a$, with $a=1/(1+z)$,  $-3.0\le w_0 \le -0.3$, and  $-2\le w_a \le 2$. 

\subsection{Results}

Results and cosmological models are listed in Table~\ref{tab_cosm_par}. By itself, time delay cosmography can pin down only the Hubble constant. According to the different priors, the central estimate of $h$ varies between 0.65 and 0.69. The related uncertainty shrinks from 0.06 in the more general scenario (OWACDM, i.e, dark energy with time dependent equation-of-state in a model of unknown curvature) to 0.04 in the more specified model of universe ($\Lambda$CDM, i.e., flat universe with a cosmological constant). 

Due to the peculiar nature of the parameter degeneracies, even further specifying the flat $\Lambda$CDM model with strong additional priors on the energy budgets does not improve the accuracy on the Hubble constant. By imposing $\Omega_\mathrm{M}=0.3$, we find $h=0.67\pm0.04$. On the other end the accuracy on $h$ significantly improves when the equation of state of the dark energy is either assumed to be known or very well constrained.

Measurements of time-delay distances are usually regarded as direct estimates of the Hubble constant. However, even if the inference of $h$ with time delays is independent of local distance ladder, it suffers slightly from the assumed cosmological model.

Our results are in good agreement with previous analyses. Recent time-delay determinations of the Hubble constant compare well with independent methods. \citet{rie+al11} employed the distance ladder and observations of type Ia supernovae to measure $h=0.738\pm0.024$. \citet{fre+al12} applied a mid-infrared calibration to the Hubble Space Telescope (HST) Key Project sample and found  $h=0.743\pm0.021(\mathrm{syst})$. The first analysis of CMB data with Planck estimated $h=0.673\pm0.012$ under the assumption of a flat $\Lambda$CDM model \citep{plaXVI}. However, Planck data alone cannot constrain $H_0$ in a generic cosmological scenario.

The most recent inference of the Hubble constant with time delays using PixeLens is from \citet{pa+hj10}, who estimated the Hubble constant to be $h=0.66^{+0.06}_{-0.04}$ (for a flat $\Lambda$CDM model with $\Omega_\mathrm{M}=0.3$). They modelled simultaneously 18 time delay lenses coupled by a shared Hubble parameter. 

To constrain also dark energy, we had to fit each lens separately and then combine the results on the time delay distances to infer the cosmological parameters. The final agreement between these two different approaches exploiting PixeLens confirms the validity of the method. 

\citet{ogu07} took an alternative approach. He derived the expected distributions of time delays by adopting realistic lens potentials and used, in turn, the distribution to derive statistically the value of the Hubble constant from observed time delays and image positions. \citet{ogu07} found that 16 published time delay quasars constrained $h$ to be $0.68\pm0.06$(stat.)$\pm0.08$(syst.) in a flat $\Lambda$CDM model with $\Omega_\mathrm{M}=0.24$.

Dark energy properties can be constrained when time-delays are complemented by analyses of the cosmic microwave background. Results in combination with the Wilkinson Microwave Anisotropy Probe nine-year (WMAP9) data are listed in Table~\ref{tab_cosm_par}. Under the assumption of spatial flatness, $w_0$ is determined to $\gs 0.15$. 

There is no evidence for physics beyond the standard $\Lambda$CDM paradigm. The equation-of-state parameter $w_0$ is highly compatible with the cosmological constant case ($w_0=-1$) in either the OWCDM or the WCDM case.

Our results agree well with those of \citet{suy+al13}, who analysed lensing, stellar kinematics and constraints on the environment for RX~J1131-1231 and B~1608+656. In combination with WMAP7 they found $h=0.75\pm0.04$, $\Omega_\mathrm{M}=0.24\pm0.03$ and $w_0=-1.14^{+0.17}_{-0.20}$ in a flat WCDM model, which are $\sim 1$ combined $\sigma$ away from our results.

\subsection{Systematics}
\label{sec_syst}

Some systematics effects might play a role in our analysis. Present samples of lensed quasars are quite heterogeneous and might be affected by selection biases. Current time delay lenses have significantly larger image separations on average compared with the other lenses, which is an indication that they likely lie in dense environments \citep{ogu07}. To estimate this effect, we repeated the analysis after excluding five clusters with image separations larger than $3\arcsec$ (RX~J0911+0551, Q~J0957+561, SDSS~J1004+4112, HE~1104-1805, RX~J1131-1231). We found $h=0.63\pm0.08$ in the WCDM model, which is not statistically separable from the result with the full sample.

Another source of systematic error is due to line of sight structures. Being the universe not homogeneous, light beams usually travels along under-dense paths and are slightly demagnified \citep{dy+ro73}. This effect is most dramatic in the so called empty beam approximation, when the light path is completely deprived of dark matter. We can estimate the size of the effect using standard formulae for the angular diameter distances in a clumpy universe with dark energy \citep{ser+al01,ser+al02}. For a typical lens configuration, i.e, deflector at $z_\mathrm{d}\sim0.5$ and source at $z_\mathrm{d}\sim2.0$, in a flat $\Lambda$CDM model with $\Omega_\mathrm{M}=0.3$, the time-delay distance can be over-estimated as much as $\gs 7$ per cent. 

This is equivalent to the effect of an external convergence of $\kappa_\mathrm{ext} \simeq -0.07$. The time delay distance measured neglecting the external contribution is the biased according to $D_{\Delta t}^\mathrm{obs} =(1-\kappa_\mathrm{ext})D_{\Delta t}^\mathrm{true}$. However, for the most cases, the effect is less severe. The distribution of $\kappa_\mathrm{ext}$ over all line of sights peaks around $-0.01$-$0.02$ \citep{ser+al02,suy+al10} and relative over-estimates of $h$ are of order of 1-2 per cent. 

Dense surroundings have an opposite effect. The external convergence due to galaxy environment with lensing bias taken into account is approximately $\kappa_\mathrm{ext}=0.03\pm0.03$ unless the image separation is too large \citep{ogu07}. The effect is much larger when the lenses, such as  RX~J0911+0551 and Q~0957+561, reside in cluster of galaxies. The external convergence was estimated to be $0.2 \ls \kappa_\mathrm{ext} \ls 0.3$  for RX~J0911+0551 \citep{hjo+al02} and $\kappa_\mathrm{ext}=0.166\pm0.056$ for Q~0957+561 \citep{nak+al09}. The effect of these two extreme cases can be quantified by repeating the analysis after rescaling the time delay distances obtained with PixeLens by $D_{\Delta t}^\mathrm{obs}/(1-\kappa_\mathrm{ext})$. We found $h=0.65\pm0.07$ for the WCDM model. 

The systematic effects discussed above are connected to some degree. Firstly, the lenses known to be in dense clusters show very large image separations.  Secondly, the effects of both the line of sight structures and the lens environment can be characterised by a single parameter, the external convergence $\kappa_\mathrm{ext}$. For under-dense line of sights $\kappa_\mathrm{ext}$ is negative whereas $\kappa_\mathrm{ext}$ is positive in dense environment or very crowded line of sights. Given the above considerations, we can then estimate the total effects of systematics to be $\delta h \ls 0.04$.

An effective approach to directly estimate the mass distribution associated with the lens galaxy and the structures along the line of sight has been already successfully employed for a couple of lenses. \citet{suy+al10} and \citet{suy+al13} estimated $\kappa_\mathrm{ext}$ via observations of the lens environment and ray tracing through numerical simulations.

\section{Conclusions}
\label{sec_conc}

Time delay cosmography is a promising tool. It can estimate the Hubble constant in a way completely independent of the local cosmic distance ladder and is sensitive to the dark energy equation-of-state too. These strengths make gravitational lens time delay nicely complementary to CMB analyses.

Parametric studies of well observed lenses can determine the time-delay distance to $\sim 6$ per cent \citep{suy+al13}. However, parametric analyses might be affected by the mass-sheet degeneracy. In fact this degeneracy is not only linked to extrinsic convergence but also to the lens density slope and might play a role even if the lensing analysis incorporates velocity dispersion measurements and a characterisation of the line of sight structures \citep{sc+sl13}.

\citet{suy+al13b} examined the issue of the systematic error introduced by an assumed lens model density profile and whether the mass-sheet degeneracy severely affects only systems with point-like images. They showed that the spatially extended Einstein ring of the lensed source and the availability of multiple time delays provide strong constraints on the local profile of the lens mass distribution. The related uncertainty over the parametric lens modelling is of the order of $\ls 3$ per cent for a system as well constrained as RX~J1131-1231, whose time delay distance can be still determined to $\sim 6$ per cent accuracy \citep{suy+al13b} .

We performed an analysis of 18 systems exploiting lensing-only information, i.e, image positions and time delays. Lenses were free-form modelled in a way that does not artificially breaks the degeneracy between the steepness of the profile and the cosmological parameters. We could determine the Hubble constant to $\ls 10$ per cent and, in combination with WMAP9, the dark energy equation-of-state to $\gs 0.15$. These uncertainties are similar to what was found parametrically in \citet{suy+al13} with the detailed analysis of only two lenses.

Parametric and free-form modelling of time-delay lenses feature complementary qualities and solve different problems. Free-form methods are unbiased for the slope-cosmology degeneracy but suffer from contamination due to line of sight structures or lens environment. The parametric approach succeeds in determining the effect of external convergence but might be still affected by the degeneracy between mass profile and cosmological parameters.

Free-form modelling is very cost effective. It can be performed without velocity dispersion measurements and does not require time-delay measurements to 1 per cent precision. On the other hand, parametric approaches can reach the same accuracy on cosmological parameter determinations with $\sim$10-20 times fewer lenses. The critical point is to properly assess the degree of mass-sheet degeneracy and the related uncertainty. The next step in the development of time delay cosmography is to combine and complement the strengths of the two approaches.

\section*{Acknowledgements}
D.~P. acknowledges financial support from Agence Nationale de la Recherche ANR-09-BLAN-0234. This research has made use of NASA's Astrophysics Data System.


\begin{thebibliography}{}
\setlength{\itemindent}{-2.5em}

\bibitem[\protect\citeauthoryear{{Coles}}{{Coles}}{2008}]{col08}
{Coles} J.,  2008, \apj, 679, 17

\bibitem[\protect\citeauthoryear{{Courbin}, {Chantry}, {Revaz}, {Sluse},
  {Faure}, {Tewes}, {Eulaers}, {Koleva}, {Asfandiyarov}, {Dye}, {Magain}, {van
  Winckel}, {Coles}, {Saha}, {Ibrahimov} \& {Meylan}}{{Courbin}
  et~al.}{2011}]{cou+al11}
{Courbin} F.,  {Chantry} V.,  {Revaz} Y.,  {Sluse} D.,  {Faure} C.,  {Tewes}
  M.,  {Eulaers} E.,  {Koleva} M.,  {Asfandiyarov} I.,  {Dye} S.,  {Magain} P.,
   {van Winckel} H.,  {Coles} J.,  {Saha} P.,  {Ibrahimov} M.,    {Meylan} G.,
  2011, \aap, 536, A53

\bibitem[\protect\citeauthoryear{{Dyer} \& {Roeder}}{{Dyer} \&
  {Roeder}}{1973}]{dy+ro73}
{Dyer} C.~C.,  {Roeder} R.~C.,  1973, \apjl, 180, L31

\bibitem[\protect\citeauthoryear{{Eulaers} \& {Magain}}{{Eulaers} \&
  {Magain}}{2011}]{eu+ma11}
{Eulaers} E.,  {Magain} P.,  2011, \aap, 536, A44

\bibitem[\protect\citeauthoryear{{Eulaers}, {Tewes}, {Magain}, {Courbin},
  {Asfandiyarov}, {Ehgamberdiev}, {Rathna Kumar}, {Stalin}, {Prabhu}, {Meylan}
  \& {Van Winckel}}{{Eulaers} et~al.}{2013}]{eul+al13}
{Eulaers} E.,  {Tewes} M.,  {Magain} P.,  {Courbin} F.,  {Asfandiyarov} I.,
  {Ehgamberdiev} S.,  {Rathna Kumar} S.,  {Stalin} C.~S.,  {Prabhu} T.~P.,
  {Meylan} G.,    {Van Winckel} H.,  2013, \aap, 553, A121

\bibitem[\protect\citeauthoryear{{Fohlmeister}, {Kochanek}, {Falco}, {Morgan}
  \& {Wambsganss}}{{Fohlmeister} et~al.}{2008}]{foh+al08}
{Fohlmeister} J.,  {Kochanek} C.~S.,  {Falco} E.~E.,  {Morgan} C.~W.,
  {Wambsganss} J.,  2008, \apj, 676, 761

\bibitem[\protect\citeauthoryear{{Fohlmeister}, {Kochanek}, {Falco},
  {Wambsganss}, {Oguri} \& {Dai}}{{Fohlmeister} et~al.}{2013}]{foh+al13}
{Fohlmeister} J.,  {Kochanek} C.~S.,  {Falco} E.~E.,  {Wambsganss} J.,  {Oguri}
  M.,    {Dai} X.,  2013, \apj, 764, 186

\bibitem[\protect\citeauthoryear{{Freedman}, {Madore}, {Scowcroft}, {Burns},
  {Monson}, {Persson}, {Seibert} \& {Rigby}}{{Freedman}
  et~al.}{2012}]{fre+al12}
{Freedman} W.~L.,  {Madore} B.~F.,  {Scowcroft} V.,  {Burns} C.,  {Monson} A.,
  {Persson} S.~E.,  {Seibert} M.,    {Rigby} J.,  2012, \apj, 758, 24

\bibitem[\protect\citeauthoryear{{Gilmore} \& {Natarajan}}{{Gilmore} \&
  {Natarajan}}{2009}]{gi+na09}
{Gilmore} J.,  {Natarajan} P.,  2009, \mnras, 396, 354

\bibitem[\protect\citeauthoryear{{Goicoechea} \& {Shalyapin}}{{Goicoechea} \&
  {Shalyapin}}{2010}]{goi+al10}
{Goicoechea} L.~J.,  {Shalyapin} V.~N.,  2010, \apj, 708, 995

\bibitem[\protect\citeauthoryear{{Goicoechea}, {Shalyapin}, {Koptelova},
  {Gil-Merino}, {Zheleznyak} \& {Ull{\'a}n}}{{Goicoechea}
  et~al.}{2008}]{goi+al08}
{Goicoechea} L.~J.,  {Shalyapin} V.~N.,  {Koptelova} E.,  {Gil-Merino} R.,
  {Zheleznyak} A.~P.,    {Ull{\'a}n} A.,  2008, \na, 13, 182

\bibitem[\protect\citeauthoryear{{Hjorth}, {Burud}, {Jaunsen}, {Schechter},
  {Kneib}, {Andersen}, {Korhonen}, {Clasen}, {Kaas}, {{\O}stensen}, {Pelt} \&
  {Pijpers}}{{Hjorth} et~al.}{2002}]{hjo+al02}
{Hjorth} J.,  {Burud} I.,  {Jaunsen} A.~O.,  {Schechter} P.~L.,  {Kneib} J.-P.,
   {Andersen} M.~I.,  {Korhonen} H.,  {Clasen} J.~W.,  {Kaas} A.~A.,
  {{\O}stensen} R.,  {Pelt} J.,    {Pijpers} F.~P.,  2002, \apjl, 572, L11

\bibitem[\protect\citeauthoryear{{Jullo}, {Natarajan}, {Kneib}, {D'Aloisio},
  {Limousin}, {Richard} \& {Schimd}}{{Jullo} et~al.}{2010}]{jul+al10}
{Jullo} E.,  {Natarajan} P.,  {Kneib} J.-P.,  {D'Aloisio} A.,  {Limousin} M.,
  {Richard} J.,    {Schimd} C.,  2010, Science, 329, 924

\bibitem[\protect\citeauthoryear{{Kochanek}}{{Kochanek}}{2004}]{koc04}
{Kochanek} C.~S.,  2004, arXiv:astro-ph/0407232

\bibitem[\protect\citeauthoryear{{Linder}}{{Linder}}{2011}]{lin11}
{Linder} E.~V.,  2011, \prd, 84, 123529

\bibitem[\protect\citeauthoryear{{Lovell}, {Jauncey}, {Reynolds}, {Wieringa},
  {King}, {Tzioumis}, {McCulloch} \& {Edwards}}{{Lovell}
  et~al.}{1998}]{lov+al98}
{Lovell} J.~E.~J.,  {Jauncey} D.~L.,  {Reynolds} J.~E.,  {Wieringa} M.~H.,
  {King} E.~A.,  {Tzioumis} A.~K.,  {McCulloch} P.~M.,    {Edwards} P.~G.,
  1998, \apjl, 508, L51

\bibitem[\protect\citeauthoryear{{Lubini}, {Sereno}, {Coles}, {Jetzer} \&
  {Saha}}{{Lubini} et~al.}{2013}]{lub+al13}
{Lubini} M.,  {Sereno} M.,  {Coles} J.,  {Jetzer} P.,    {Saha} P.,  2013,
  MNRAS submitted

\bibitem[\protect\citeauthoryear{{Nakajima}, {Bernstein}, {Fadely}, {Keeton} \&
  {Schrabback}}{{Nakajima} et~al.}{2009}]{nak+al09}
{Nakajima} R.,  {Bernstein} G.~M.,  {Fadely} R.,  {Keeton} C.~R.,
  {Schrabback} T.,  2009, \apj, 697, 1793

\bibitem[\protect\citeauthoryear{{Oguri}}{{Oguri}}{2007}]{ogu07}
{Oguri} M.,  2007, \apj, 660, 1

\bibitem[\protect\citeauthoryear{{Oguri}, {Ofek}, {Inada}, {Morokuma}, {Falco},
  {Kochanek}, {Kayo}, {Broadhurst} \& {Richards}}{{Oguri}
  et~al.}{2008}]{ogu+al08}
{Oguri} M.,  {Ofek} E.~O.,  {Inada} N.,  {Morokuma} T.,  {Falco} E.~E.,
  {Kochanek} C.~S.,  {Kayo} I.,  {Broadhurst} T.,    {Richards} G.~T.,  2008,
  \apjl, 676, L1

\bibitem[\protect\citeauthoryear{{Oscoz}, {Alcalde}, {Serra-Ricart},
  {Mediavilla}, {Abajas}, {Barrena}, {Licandro}, {Motta} \&
  {Mu{\~n}oz}}{{Oscoz} et~al.}{2001}]{osc+al01}
{Oscoz} A.,  {Alcalde} D.,  {Serra-Ricart} M.,  {Mediavilla} E.,  {Abajas} C.,
  {Barrena} R.,  {Licandro} J.,  {Motta} V.,    {Mu{\~n}oz} J.~A.,  2001, \apj,
  552, 81

\bibitem[\protect\citeauthoryear{{Paraficz} \& {Hjorth}}{{Paraficz} \&
  {Hjorth}}{2010}]{pa+hj10}
{Paraficz} D.,  {Hjorth} J.,  2010, \apj, 712, 1378

\bibitem[\protect\citeauthoryear{{Planck Collaboration}, {Ade}, {Aghanim},
  {Armitage-Caplan}, {Arnaud}, {Ashdown}, {Atrio-Barandela}, {Aumont},
  {Baccigalupi}, {Banday} \& et al.}{{Planck Collaboration}
  et~al.}{2013}]{plaXVI}
{Planck Collaboration} {Ade} P.~A.~R.,  {Aghanim} N.,  {Armitage-Caplan} C.,
  {Arnaud} M.,  {Ashdown} M.,  {Atrio-Barandela} F.,  {Aumont} J.,
  {Baccigalupi} C.,  {Banday} A.~J.,    et al. 2013, arXiv:1303.5076

\bibitem[\protect\citeauthoryear{{Refsdal}}{{Refsdal}}{1964}]{ref64}
{Refsdal} S.,  1964, \mnras, 128, 307

\bibitem[\protect\citeauthoryear{{Riess}, {Macri}, {Casertano}, {Lampeitl},
  {Ferguson}, {Filippenko}, {Jha}, {Li} \& {Chornock}}{{Riess}
  et~al.}{2011}]{rie+al11}
{Riess} A.~G.,  {Macri} L.,  {Casertano} S.,  {Lampeitl} H.,  {Ferguson} H.~C.,
   {Filippenko} A.~V.,  {Jha} S.~W.,  {Li} W.,    {Chornock} R.,  2011, \apj,
  730, 119

\bibitem[\protect\citeauthoryear{{Saha}, {Coles}, {Macci{\`o}} \&
  {Williams}}{{Saha} et~al.}{2006}]{sah+al06}
{Saha} P.,  {Coles} J.,  {Macci{\`o}} A.~V.,    {Williams} L.~L.~R.,  2006,
  \apjl, 650, L17

\bibitem[\protect\citeauthoryear{{Saha} \& {Williams}}{{Saha} \&
  {Williams}}{2004}]{sa+wi04}
{Saha} P.,  {Williams} L.~L.~R.,  2004, \aj, 127, 2604

\bibitem[\protect\citeauthoryear{{Schneider} \& {Sluse}}{{Schneider} \&
  {Sluse}}{2013}]{sc+sl13}
{Schneider} P.,  {Sluse} D.,  2013, ArXiv:1306.0901

\bibitem[\protect\citeauthoryear{{Sereno}}{{Sereno}}{2002}]{ser02}
{Sereno} M.,  2002, \aap, 393, 757

\bibitem[\protect\citeauthoryear{{Sereno}, {Covone}, {Piedipalumbo} \& {de
  Ritis}}{{Sereno} et~al.}{2001}]{ser+al01}
{Sereno} M.,  {Covone} G.,  {Piedipalumbo} E.,    {de Ritis} R.,  2001, \mnras,
  327, 517

\bibitem[\protect\citeauthoryear{{Sereno} \& {Longo}}{{Sereno} \&
  {Longo}}{2004}]{se+lo04}
{Sereno} M.,  {Longo} G.,  2004, \mnras, 354, 1255

\bibitem[\protect\citeauthoryear{{Sereno}, {Piedipalumbo} \& {Sazhin}}{{Sereno}
  et~al.}{2002}]{ser+al02}
{Sereno} M.,  {Piedipalumbo} E.,    {Sazhin} M.~V.,  2002, \mnras, 335, 1061

\bibitem[\protect\citeauthoryear{{Sereno} \& {Zitrin}}{{Sereno} \&
  {Zitrin}}{2012}]{se+zi12}
{Sereno} M.,  {Zitrin} A.,  2012, \mnras, 419, 3280

\bibitem[\protect\citeauthoryear{{Soucail}, {Kneib} \& {Golse}}{{Soucail}
  et~al.}{2004}]{sou+al04}
{Soucail} G.,  {Kneib} J.-P.,    {Golse} G.,  2004, \aap, 417, L33

\bibitem[\protect\citeauthoryear{{Suyu}, {Auger}, {Hilbert}, {Marshall},
  {Tewes}, {Treu}, {Fassnacht}, {Koopmans}, {Sluse}, {Blandford}, {Courbin} \&
  {Meylan}}{{Suyu} et~al.}{2013}]{suy+al13}
{Suyu} S.~H.,  {Auger} M.~W.,  {Hilbert} S.,  {Marshall} P.~J.,  {Tewes} M.,
  {Treu} T.,  {Fassnacht} C.~D.,  {Koopmans} L.~V.~E.,  {Sluse} D.,
  {Blandford} R.~D.,  {Courbin} F.,    {Meylan} G.,  2013, \apj, 766, 70

\bibitem[\protect\citeauthoryear{{Suyu}, {Marshall}, {Auger}, {Hilbert},
  {Blandford}, {Koopmans}, {Fassnacht} \& {Treu}}{{Suyu}
  et~al.}{2010}]{suy+al10}
{Suyu} S.~H.,  {Marshall} P.~J.,  {Auger} M.~W.,  {Hilbert} S.,  {Blandford}
  R.~D.,  {Koopmans} L.~V.~E.,  {Fassnacht} C.~D.,    {Treu} T.,  2010, \apj,
  711, 201

\bibitem[\protect\citeauthoryear{{Suyu}, {Treu}, {Hilbert}, {Sonnenfeld},
  {Auger}, {Blandford}, {Collett}, {Courbin}, {Fassnacht}, {Koopmans},
  {Marshall}, {Meylan}, {Spiniello} \& {Tewes}}{{Suyu}
  et~al.}{2013b}]{suy+al13b}
{Suyu} S.~H.,  {Treu} T.,  {Hilbert} S.,  {Sonnenfeld} A.,  {Auger} M.~W.,
  {Blandford} R.~D.,  {Collett} T.,  {Courbin} F.,  {Fassnacht} C.~D.,
  {Koopmans} L.~V.~E.,  {Marshall} P.~J.,  {Meylan} G.,  {Spiniello} C.,
  {Tewes} M.,  2013b, arXiv:1306.4732

\bibitem[\protect\citeauthoryear{{Tewes}, {Courbin}, {Meylan}, {Kochanek},
  {Eulaers}, {Cantale}, {Mosquera}, {Magain}, {Van Winckel}, {Sluse},
  {Cataldi}, {Voros} \& {Dye}}{{Tewes} et~al.}{2012}]{tew+al12}
{Tewes} M.,  {Courbin} F.,  {Meylan} G.,  {Kochanek} C.~S.,  {Eulaers} E.,
  {Cantale} N.,  {Mosquera} A.~M.,  {Magain} P.,  {Van Winckel} H.,  {Sluse}
  D.,  {Cataldi} G.,  {Voros} D.,    {Dye} S.,  2012, arXiv:1208.6009

\bibitem[\protect\citeauthoryear{{Treu}, {Marshall}, {Cyr-Racine}, {Fassnacht},
  {Keeton}, {Linder}, {Moustakas}, {Bradac}, {Buckley-Geer}, {Collett},
  {Courbin}, {Dobler} \& {et al.}}{{Treu} et~al.}{2013}]{tre+al13}
{Treu} T.,  {Marshall} P.~J.,  {Cyr-Racine} F.-Y.,  {Fassnacht} C.~D.,
  {Keeton} C.~R.,  {Linder} E.~V.,  {Moustakas} L.~A.,  {Bradac} M.,
  {Buckley-Geer} E.,  {Collett} T.,  {Courbin} F.,  {Dobler} G.,    {et al.}
  2013, arXiv:1306.1272

\bibitem[\protect\citeauthoryear{{Ull{\'a}n}, {Goicoechea}, {Zheleznyak},
  {Koptelova}, {Bruevich}, {Akhunov} \& {Burkhonov}}{{Ull{\'a}n}
  et~al.}{2006}]{ull+al06}
{Ull{\'a}n} A.,  {Goicoechea} L.~J.,  {Zheleznyak} A.~P.,  {Koptelova} E.,
  {Bruevich} V.~V.,  {Akhunov} T.,    {Burkhonov} O.,  2006, \aap, 452, 25

\bibitem[\protect\citeauthoryear{{Vuissoz}, {Courbin}, {Sluse}, {Meylan},
  {Chantry}, {Eulaers}, {Morgan}, {Eyler}, {Kochanek}, {Coles}, {Saha},
  {Magain} \& {Falco}}{{Vuissoz} et~al.}{2008}]{vui+al08}
{Vuissoz} C.,  {Courbin} F.,  {Sluse} D.,  {Meylan} G.,  {Chantry} V.,
  {Eulaers} E.,  {Morgan} C.,  {Eyler} M.~E.,  {Kochanek} C.~S.,  {Coles} J.,
  {Saha} P.,  {Magain} P.,    {Falco} E.~E.,  2008, \aap, 488, 481

\bibitem[\protect\citeauthoryear{{Vuissoz}, {Courbin}, {Sluse}, {Meylan},
  {Ibrahimov}, {Asfandiyarov}, {Stoops}, {Eigenbrod}, {Le Guillou}, {van
  Winckel} \& {Magain}}{{Vuissoz} et~al.}{2007}]{vui+al07}
{Vuissoz} C.,  {Courbin} F.,  {Sluse} D.,  {Meylan} G.,  {Ibrahimov} M.,
  {Asfandiyarov} I.,  {Stoops} E.,  {Eigenbrod} A.,  {Le Guillou} L.,  {van
  Winckel} H.,    {Magain} P.,  2007, \aap, 464, 845

\end{thebibliography}

\setlength{\bibhang}{2.0em}

\end{document}